# DEVELOPMENT OF A MULTIFACTOR AUTHENTICATION RESULT CHECKER SYSTEM THROUGH GSM

Famutimi Rantiola[1], Emuoyibofarhe Ozichi[2], AkinpuleAbiodun[3], Gambo Ishaya[4],Odeleye Damilola[5]

[1,2,3,5]Computer Science & Info. Technology Dept., Bowen University, Nigeria

[4]Computer Science & Engineering Dept. Obafemi Awolowo University, Nigeria

## ABSTRACT

*This work is an implementation of a multifactor authentication SMS based result checking system. The objectives of this work were to improve on the available authentication methods and apply it on examination result checking system. The work takes care of only course codes with their grades, the current GPA and the overall CGPA. It employs the Pull SMS service, built on an independent service and a modem. Examination results consist of sensitive information, hence the need to further enhance the ones already in place so as to ensure further privacy and integrity.In the course of this project, the following assumptions were made: That a system that does the computation of students' result, calculation of GPA and CGPA is already in place. The implemented system was connected to the database of the existing system. A database that contains the bio-data of each student admitted exists. That SMTP(Simple Mail Transport Protocol) modem exists and should have been used but to reduce cost, a modem that can act like a SIM browser is used with a standard SIM card inserted in it and connected via cable to the application server. The system showed that further security and privacy could be achieved when multifactor authentication is employed. For further work, the system could be developed and built as a Dependent Service which involves having the application server connected to the service provider's SMS Centre (SMSC).*

## 1. INTRODUCTION

For many years now, computers and Global System for Mobile Communications (GSM) devices have been widely used and have been a great medium of communication. With the advent of the internet and intranet, communication between people have been widely improved and made easier. The internet is a global system of computer networks that use the standard internet protocol suite to serve billions of users worldwide. In this era of modern technology, the role of telecommunication cannot also be ignored as it has a crucial impact in making communication easier. Telecommunication can be referred to as the science and practice of transmitting information by electromagnetic means. Communication is talking to someone or thing, not necessarily t hrough technology. However, Telecommunication means "talking through technology".

## 2. SMS

SMS is an acronym for Short Messaging Services. It is a mobile technology that allows for short text messages to be sent from one cell phone to another cell phone or from the Web to another cell phone using standardized communication protocols [1]. By communication protocols, we mean a system of digital message formats and rules for exchanging those messages in or between computing systems and in telecommunication. SMS messages often use a T9 predictive technology which makes text messaging faster and more efficient on non-QWERTY cell phones without full keyboards. The major advantage of text messaging is availability and cost effectiveness.

### 2.1. Application areas of SMS

The use of SMS has gone into many disciplines and professions. The worldwide burden of chronic disease is widespread and growing. This shift from acute to chronic care requires rethinking how resources are invested in managing these conditions[2]. One response has been to create programs and interventions that have the goal of helping patients better manage their own conditions. Over time, these self-management interventions and strategies have increasingly relied on various technologies for their implementation, with the newest technology being mobile phones and short message service (SMS), In medicine practice SMS is now used to monitor patients

### 2.2. Multifactor Authentication System

Authentication as the process used to determine whether the person or process attempting to gain access to a secure system is the person he or she claims to be[3]. Additionally, multi factor authentication was defined as a security authentication system in which more than one form of authentication is used to validate the identity of a user. It specifies various authentication factors that can be combined in a multifactor authentication system and this include; Knowledge- based questions, Grid card challenge

### 2.3. What is a mobile phone?

A mobile phone (also as a cellular phone, cell phone and hand phone) is a device that can make and receive telephone calls over a radio link while moving around a wide geographical area. It does so by connecting to a cellular network provided by a mobile phone operator, allowing access to the public telephone network. Other Services incorporated include short messaging services (SMS), Multimedia Messaging (MMS), chat rooms, blackberry messaging (BBM), emails, internet access, short-range wireless communications (infrared, Bluetooth), business applications, gaming and photography. The simplest and most generic of these is the SMS services. All mobile phones these days, even the cheapest have the functionality to send and receive messages.

## 3. STATEMENT OF PROBLEM

In many countries of the world today, there still exists a problem of safe checking of examination results. Most students are always anxious   when results are released / pasted, there is usually panic here and there. A whole lot of students throng to the notice board to know their fate. Some even get injured in the process. Although, in many universities today (using Bowen University

Nigeria as a case study), a university portal where students can access their results is already in place but there can also be a simpler way of getting this done instead of getting angry or frustrated about slow or bad network connectivity during the course of checking online. Also, the availability of internet facilities in many countries is low and quite expensive. In this regard, a system was designed to make result checking faster, simpler and less hazardous exercise. Some organizations that are already using this type of system are not considering the need to enforce proper privacy of students through   enhanced authentication systems.

## 3.1. Aim and objectives

The aim of this research is to implement a multifactor authentication SMS based result checking system.  The Objectives are to: (i) to design a model for a multifactor authentication SMS result check system (ii) to implement the model and (iii) to test the system.

## 3.2. Scope of study

The research takes care of only course codes with their grades, and the overall GPA and CGPA. It does not show the marks of the students. It only employs the Pull SMS service, built on an independent service. An independent service involves using an application server (i.e. the system running the application) and a modem. This service offers limited benefit but it is easy and fast to set up because it does not require authorization of the service provider (mobile network operators) or connection to other third party SMS provider. Themodem uses a SIM card which has a normal phone number and messages that originate from the modem attracts the standard cost or tariff.However, in the course of this project, several assumptions were made and these are listed as follows (i) That a system that does the computation of students' result, calculation of GPA and CGPA is already in place. The implemented system was connected to the database of the existing system. (ii) A database that contains the bio-data of each student admitted exists. (iii) That SMTP(Simple Mail Transport Protocol) modem exists and should have been used but to reduce cost, a modem that can act like a SIM browser is used with a standard SIM card inserted in it and connected via cable to the application server.

# 4. REVIEW OF RELATED WORKS

Some related works have been studied. Those that are very relevant have been explained below.

## 4.1. (NTU eXpress SMS)

This system is used at the Nanyang Technological University. It mandates undergraduate students to register their mobile phones on the NeXS Portal. These phones can then be used to send SMS and access information. Unregistered mobile numbers are not allowed by the system and only three mobile providers are acceptable (M1/Singtel/Starhub).For example, to get exam results via NeXS, users send the keyword NTU RESULT to 74000.

### 4.1.1. Strength of NeXS

The system provides some sort of security by allowing only pre-registered mobile numbers to have access. Other SMS services like seating arrangements, subject time table and library account information are included.

### 4.1.2. Weakness of NeXS

The system is not flexible enough as a constraint has been placed on   access to the system. It also poses a threat of identifying if it's the actual owner of the phone that sent the request. Anybody can pick the   phone, send the SMS to the short code and get the required result. This invades the privacy of the real owner of the phone. Also, in the event of carelessness or theft whereby the mobile phone gets lost, the user would not have access to any information unless he/she visits the website and re-register a new phone. The system also does not give students access to past results, only newly-concluded semester results are made available.

## 4.2. Malaysian University English Test (MUET) SMS Result Checking System

This system receives in the following format:MUET <space> IC Number to 39003.The charge rate is 15 cents for each message sent and 30 cents for each received.

### 4.2.1. Weaknesses of MUET

The system does not offer any form of security because it uses only the roll number. So, anybody can access another person's information, provided he/she knows the IC number. Only current semester results can be accessed, old ones are not accessible. It is also somewhat expensive on the side of students as tariffs on received messages are higher than that of sent messages, hence, discouraging the students from using the system.

## 4.3. SMS Result Checking of Board of Secondary Education, Orissa

The school at Orissa makes available the Higher School Certificate (HSC) exam result available via SMS. The examination result is requested by sending the SMS Code: HSCR RollNo (for regular/regular Correspondence courses) and HSCX Roll-No (for Ex-regular Correspondence Courses) to 56505 [4].

### 4.3.1. Weakness of the System

The system allows request only from BSNL Mobile and does not provide any form of security. Students' result can then be accessed by specifying the roll number.

## 4.4. An improved SMS interface result checking system

Animproved SMS interface result checking was developed to address some of the problems identified with some available systems [5, 6].

### 4.4.1. Strength of the System

The system assigns a default password to each student. This password is expected to be changed on the school's website or bysending SMS to this effect to the university. Specifying a password in the SMS ensures that if a cell phone gets stolen or get into the wrong hands, thestudent's result cannot be accessed unless the password is specified. This is animprovement on the NTU eXpress SMS system. It also includes some sort of error handling to check for errors in the message format.

### 4.4.2. Shortcomings of the system

The system makes use of a form of social interaction and an additional module for password generation is created. The user sends the surname and matriculation number to a phone number, connected to the application and the system responds with the system-generated password. The user then logs in with that password and makes his request. The system is not considered as very secure because it affords any other person to access the system provided he knows the user's name and matriculation number. Also, the system generates password for each result check, introducing additional overhead. This can be very cumbersome. Also, a Sony Ericson phone with a standard SIM card is connected to the application server to fetch results from it. Here, a serial cable with one end connecting to the phone and the other end connected to the computer hosting the application is used. This is quite unreliable as either connections might be altered, causing the application to be made unavailable when this occurs. Also, at one point or the other, there would be need to charge the phone's battery so that messages can keep coming in, processed and requests sent.

## 5. OUR PROPOSED SYSTEM

The system was developed using C# language of Microsoft Visual Studio .NET framework 4.0, MS-Server 2005(Server Management Studio) and a database connector – open database connectivity (ODBC) to connect the database [7].The system has two - tier architecture comprising of the following: (i)The business-logic tier: This tier serves as the middleware that is responsible for processing the user's requests. The MS-SQL server would be employed for all business-logic processing and request. The language employed in this tier is c# language. (ii)The database tier: This serves as the repository containing various tables of students' details (names, matriculation numbers, departments, grades, GPA, CGPA). It also stores message logs, audit logs. MS-SQL database server (2005) was used as a repository for all data and information needed in this module as well as ensuring dynamism. The language employed in this tier is the structured query language (SQL).

### 5.1. Methodology

The system uses the short messaging service to provide a means of fast communication between the students and the university. It makes use of abetter social interaction mechanism. The factors used for authentication were knowledge based questions. The system mandates user to go through a combination of questions and requires that the questions be answered correctly in order to gain access to the system. The answers to these questions, however, would have been entered into the database by theadministrator as part of the student bio-data after admission into the university.To get grades:G <space> Matriculation number <space> Semester <space> Session. e.g.        G SSE/010/7600 1 2012/2013; To get CGPA and / or GPA:C <space> Matriculation number <space> Semester <space> Session. e.g. C SSE/010/7600 1 2012/2013        There is also a help option:H<space> Matriculation number <space> Semester <space> Session. e.g.H SSE/010/7600 1 2012/2013

### 5.2. Results

The illustrations and the following screen shots below show the four-leg transaction:Figure (1) shows a student requesting for his grades, Figure (2) shows the security questions sent back by the application, Figure (3) shows the correct answers provided by the student and Figure (4) shows the grades sent back by the application according to the request sent.

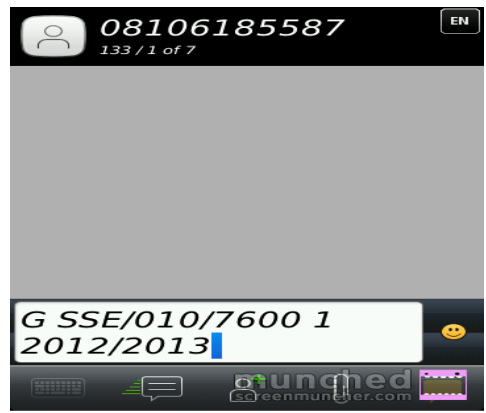

Figure 1. Request for examination grades

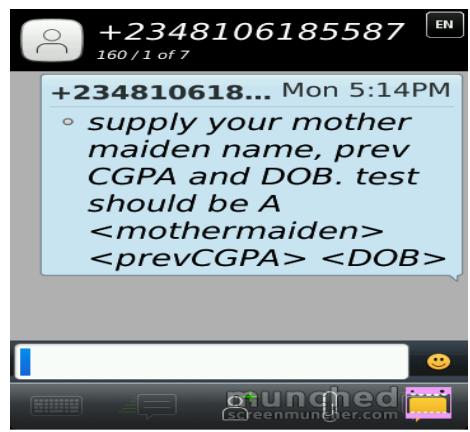

Figure 2.Security questions sent back by the application

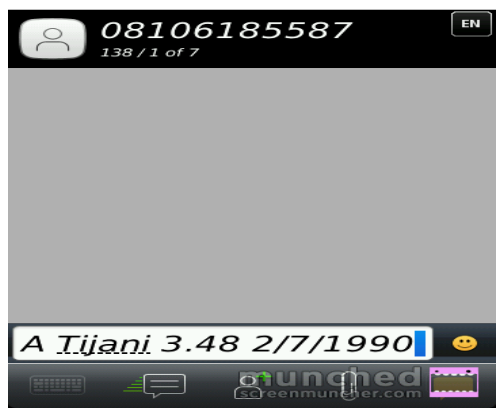

Figure 3. A valid answer provided by the student

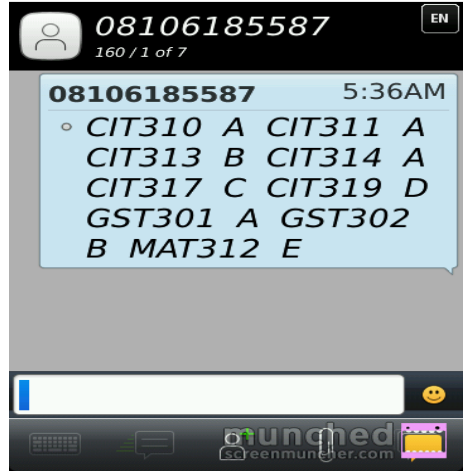

Figure 4.Grades sent back by the application according to the request sent.

# 6. OUR PROPOSED SYSTEM

The system provides a more secured and efficient way of getting results. Results are made available for all registered students of the university that have passed the authentication test. The system uses a multi-factor authentication approach by asking security based questions and awaits correct answers to allow access.In comparison with previously existing systems, this system is dynamic and much more security-conscious than the use of PINs with passwords, IDs and registered phone numbers have been completely eliminated. Hence, an open system based on trust is introduced.

# 7. RECOMMENDATION

For further work, the system could be developed and built as a Dependent Service: This involves having the application server connect to the service provider's SMS Centre (SMSC). It requires a constant connection to the internet as the application server does not require any physical phone/modem with a SIM card connected to it; rather it connects to a SMSC. When users send their request, it goes to the SMSC, which automatically forwards the message to the application server over the internet. These servers are assigned short numbers instead of the traditional 10 or 11 digits numbers. These numbers, also referred to as short codes, are usually 4 to 6 digits long and are operator specific. Also, a premium fee (a fee other than the fixed rate for SMS) can be charged on these codes.

## ACKNOWLEDGEMENTS

The authors thank all members of staff of Computer Science and Information Technology Department of Bowen University, as well as all members of Health Information Systems (HIS) research group of Obafemi Awolowo University Nigeria.

## Authors

Famutimi Rantiola holds B.Sc., M.Sc and M.Phil. degrees in Computer Science. He is
working on his PhD at Obafemi Awolowo University. He teaches ComputerScience
coursesat Bowen University, Nigeria 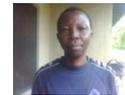

Emuoyibofarhe OzichiholdsB.Sc., M.Tech and PhD degrees in Computer Science.She currently teaches
Computer science courses at Bowen University.

Akinpelu Abiodunholds M.Tech in Computer Science and currently working on his PhD programme at
Ladoke Akintola University of Technology Nigeria. He is an Assistant Chief Technologist with Bowen
University

GamboIshayaholds B.Sc.and M.Sc degrees in ComputerScience. He is currently workingon hisPhD at
Obafemi Awolowo UniversityNigeria. He teaches Computer science courses atObafemi Awolowo
University.

Odeleye Damilolaholds B.Sc.degree in ComputerScience. She is currently working on  herM.Sc degree.